\documentclass[
showpacs,floatfix,
aps,prb,amsmath,
twocolumn,
superscriptaddress
]{revtex4}

\usepackage{epsfig}
\usepackage{epsf}
\usepackage{array}
\usepackage{bm}

\begin{document}
\newcommand{\be}{\begin{equation}}
\newcommand{\ee}{\end{equation}}
\newcommand{\bea}{\begin{eqnarray}}
\newcommand{\eea}{\end{eqnarray}}
\newcommand{\br}{{\bm r}}
\newcommand{\bk}{{\bm k}}
\newcommand{\bq}{{\bm q}}
\newcommand{\bn}{{\bm n}}
\newcommand{\bp}{{\bm p}}
\newcommand{\bE}{{\bm E}}
\newcommand{\ve}{\varepsilon}

\title{Compressibility of a 2D electron gas under microwave radiation}
\author{ M.G.~Vavilov}\altaffiliation[Present address: ]
{Department of Applied Physics, Yale University, New Haven, CT  06520}
\affiliation{Center for Materials Sciences and Engineering,
  Massachusetts Institute of Technology, Cambridge, MA 02139}

\author{I.A.~Dmitriev}\altaffiliation[Also at ]
{A.F. Ioffe Physico-Technical Institute, 194021 St.~Petersburg, Russia.}
\affiliation{Institut f\"ur Nanotechnologie, Forschungszentrum
Karlsruhe, 76021 Karlsruhe, Germany}

\author{I.L.~Aleiner}\affiliation{Physics Department, Columbia University, New York, NY
10027, USA}

\author{A.D.~Mirlin}
\altaffiliation[Also at ]{Petersburg Nuclear Physics
Institute, 188350 St.~Petersburg, Russia.}
\affiliation{Institut f\"ur Nanotechnologie, Forschungszentrum
Karlsruhe, 76021 Karlsruhe, Germany}
\affiliation{Institut f\"ur Theorie der Kondensierten Materie,
Universit\"at Karlsruhe, 76128 Karlsruhe, Germany}

\author{D.G.~Polyakov}\altaffiliation[Also at ]
{A.F. Ioffe Physico-Technical Institute, 194021 St.~Petersburg, Russia.}
\affiliation{Institut f\"ur Nanotechnologie, Forschungszentrum
Karlsruhe, 76021 Karlsruhe, Germany}
\hspace*{-1mm}
\date{September 27, 2004}

\begin{abstract}
Microwave irradiation of a two-dimensional electron gas (2DEG) produces a
non-equilibrium distribution of electrons, and leads to oscillations in the
dissipative part of the conductivity. We show that the same non-equilibrium
electron distribution induces strong oscillations in the 2DEG compressibility
measured by local probes. Local measurements of the compressibility are
expected to provide information about the domain structure of the zero
resistance state of a 2DEG under microwave radiation.
\end{abstract}

\maketitle

Microwave irradiation of a high-mobility two-dimensional electron
system (2DEG) in GaAs/AlGaAs heterostructures results in oscillations
of the dissipative part of the resistivity as a function of
$\omega/\omega_{\rm c}$, where $\omega$ is the microwave frequency and
$\omega_{\rm c}=eB/mc$ is the cyclotron frequency in magnetic field
$B$.\cite{zudov01} Remarkably, in systems with a very high mobility at
sufficiently high radiation power the amplitude of the oscillations
becomes larger than the dark value of the resistivity and regions of
zero resistance develop.\cite{mani02,zudov03} Experiments
\cite{yang03,willett03} have shown that the main features of the
effect of microwave radiation on the transport properties of a 2DEG
are independent of the sample geometry.

Theoretical studies of this phenomenon have been focused on the effect
of microwave radiation on the electron
spectrum\cite{ryzhii,durst03,vavilov03} and on the electron
distribution function.\cite{dmitriev03,DVAMP} As we have shown in
Ref.~\onlinecite{DVAMP}, it is the latter effect that yields the
leading microwave-induced contribution to the dissipative resistivity
$\rho_{xx}$, explaining its $\omega/\omega_{\rm c}$--oscillations
observed in the experiments.

The question we address in this communication is how the
non-equilibrium electron distribution, responsible for the effect of
microwaves on the longitudinal resistivity, affects the thermodynamic
properties of the 2DEG, such as the compressibility $\chi$. We find
that $\chi$ exhibits $\omega/\omega_{\rm c}$ oscillations similar to
those of the dissipative resistivity.  We demonstrate that local
measurements of the compressibility may be used as a probe of the
spatial structure of the zero resistance state; hence, such measurements
may verify
the existence of domains suggested in Ref.~\onlinecite{andreev03}.

The compressibility $\chi$ is defined as the static limit of the
density-density response function $\Pi(\omega,q)$, i.e.,
$\chi=\Pi(0,q)$ is given by $\delta n_q/e\phi_q$, where $\delta
n_q$ is the electron density induced as a linear response to the
application of a static self-consistent (screened) potential
$\phi_q$ and $-e<0$ is the electron charge. Experimentally, the
compressibility of a 2DEG in the limit of small $q$ can be
determined by measuring the capacitance between the 2DEG and a
gate \cite{smith85} or the electric field screening in
double-layer systems.\cite{eisenstein92} On the other hand, a
technique utilizing single-electron transistors has been developed
\cite{ilani00} that allows one to measure the {\it local}
compressibility at $q\sim L^{-1}$, where $L$ is the spatial scale
on which the screened potential $\phi ({\bm r})$ created by a
local probe and the induced electron density $\delta n({\bm r})$
change in the 2DEG plane.

Let us start by considering the local compressibility
$\chi$ in the limit $ql_{\rm in}\gg 1$, where $l_{\rm in}= (D_{\rm
c}\tau_{\rm in})^{1/2}$ is the inelastic relaxation length,
$\tau_{\rm in}$ is the electron-electron scattering time
calculated in Ref.~\onlinecite{DVAMP}, and $D_{\rm c}$ is the
diffusion coefficient. As shown at the end of the paper, in this
limit we can neglect the inelastic-scattering induced damping of
the $\omega/\omega_c$ oscillations of $\chi$. For a
non-equilibrium state characterized by an electron distribution
function $f(\ve)$, we then have
\be
\chi
= -\int \nu(\ve)
\frac{\partial f(\ve)}{\partial \ve} d\ve,
\label{compressGeneral}
\ee
where $\nu(\ve)$ is the density of states. In the equilibrium
case, Eq.~(\ref{compressGeneral}) reduces to the conventional
formula $\chi_0=\partial n_e/\partial\mu$, where $n_e$ is the
electron density and $\mu$ is the chemical potential.

Microwave radiation changes both the electron distribution
function $f(\ve)$ and the oscillatory component of the
electron density of states $\nu(\ve)$.
The microwave-induced correction to the compressibility $\delta\chi$
can therefore be split into three parts: (i) related to the change of the
density of states only, (ii) related to the effect of microwaves on the
distribution function only, and (iii) mixed term. We notice first of all
that the contribution (i), which is the only one that remains if the influence
of radiation on the distribution function is ignored, is exponentially
suppressed\cite{note-sigma} at $2\pi^2 T\gg\omega_c$
because of the temperature smearing, see Eq.~(\ref{compressGeneral}).
We are thus left with the contributions (ii), (iii) to the compressibility
$\chi= \chi_0+\delta\chi$, which are governed by the deviation
of the electron distribution function $f_{\rm mw}(\ve)$
from the equilibrium Fermi function $f_0(\ve)$:
\be
\delta \chi= -\int \nu_{\rm mw}(\ve)\frac{\partial }{\partial
\ve}\left[
f_{\rm mw}(\ve)-f_0(\ve )
\right] d\ve,
\label{pcomp}
\ee
where $\nu_{\rm mw}(\ve)$ is the density of states in the
presence of microwaves. Following the arguments of Ref.~[\onlinecite{DVAMP}],
the contribution (iii) can be neglected as compared to (ii) at
$\tau_{\rm in}\gg\tau_{\rm q}$, where $\tau_{\rm in}$ and
$\tau_{\rm q}$ are the inelastic scattering time and
the quantum scattering time, respectively.
In other words, we can replace $\nu_{\rm mw}(\ve)$ in Eq.~(\ref{pcomp}) by
the dark density of states $\nu(\ve)$.

The distribution function
$f_{\rm mw}$ is the solution to the following kinetic
equation:\cite{DVAMP}
\begin{equation}
\begin{split}
\label{kineq}
&\frac{{\cal P}}{4\nu_0}
\sum\limits_{\pm}\nu(\ve\pm\omega)\,
[\,f_{\rm mw}(\ve\pm\omega)-f_{\rm mw}(\ve)\,]  \\
&+\frac{{\cal Q}}{4\nu_0 {\nu}(\ve)}
\frac{\omega_{\rm c}^2 }{\pi^2}\frac{\partial}{\partial\ve}
\left[\nu^2(\ve)\frac{\partial f_{\rm mw}(\ve)}{\partial\ve}\,\right]
=f_{\rm mw}(\ve)-f_0(\ve),
\end{split}
\end{equation}
written for  $|\omega\pm\omega_{\rm c}| {\tau_{\rm tr}}\gg 1$ in terms of the
dimensionless  parameters  ${\cal P}$ and  ${\cal Q}$ characterizing
the power of the microwave field ${\cal E}_\omega$
and the strength of the dc electric field ${\cal E}_{\rm dc}$,
\begin{equation}
{\cal P}
=\frac{\tau_{\rm in}}{\tau_{\rm tr}}
\frac{e^2 {\cal E}_\omega^2 v_F^2}{\omega^2}
\frac{\omega_{\rm c}^2+\omega^2}{(\omega^2-\omega_{\rm c}^2)^2},\ \
{\cal Q}=\frac{2\tau_{\rm in}}{\tau_{\rm tr}}
\frac{\pi^2 e^2 {\cal E}_{\rm dc}^2 v_F^2}{\omega_{\rm c}^4},
\label{PQ}
\end{equation}
where $v_{\rm F}$ is the Fermi velocity, $\tau_{\rm tr}$ is the
transport (momentum relaxation) time, and $\nu_0$ is the density of states
at zero magnetic field. The first term in the left-hand
side of the kinetic equation describes processes of absorption and
emission of microwave quanta and the second term describes the
diffusion of electrons over the energy spectrum due to the constant
electric field. The right-hand side of Eq.~(\ref{kineq}) represents
the inelastic relaxation of the electron distribution function towards
the equilibrium function $f_0(\ve)$.

\begin{figure}
\epsfxsize=0.45\textwidth
\epsfbox{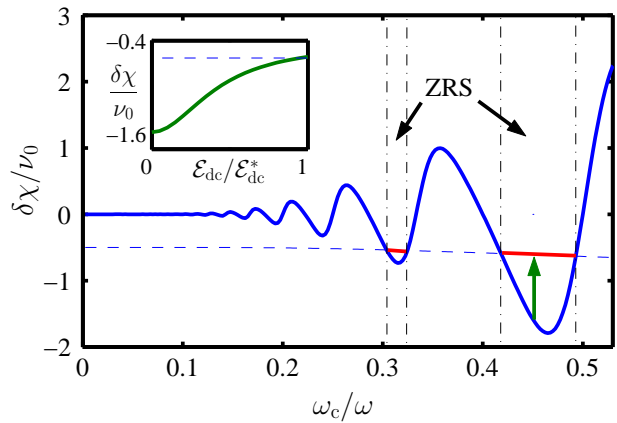}
\caption{The microwave-induced correction to the compressibility (solid line)
of a 2DEG as a function of $\omega_{\rm
c}/\omega$ at fixed $\omega\tau_{\rm q}=2\pi$ and microwave power
${\cal P}|_{\omega_c=0}=1$. In the zero resistance state (ZRS), the
electric field inside domains ${\cal E}_{\rm dc}^*$ fixes the
compressibility at the level of $\delta\chi^*$ shown by a dashed line [see
Eq.~(\ref{comprZRS})]. Inside the domain wall, the electric field ${\cal
E}_{\rm dc}$ is smaller than ${\cal E}_{\rm dc}^*$ and the compressibility
depends on the local field as shown in the inset (for
$\omega_{\rm c}/\omega=0.45$, this ratio is indicated by the arrow).
}
\label{fig:q1}
\end{figure}

The local relaxation-time approximation for inelastic collisions is justified
in the calculation of the local compressibility by means of Eq.~(\ref{pcomp})
if $q_1L\gg 1$, where $q_1=(\omega^3_{\rm c}\tau_{\rm tr})^{1/2}/v_{\rm F}$
is the characteristic momentum
transfer in the electron-electron collision integral.\cite{q_1} Under this
condition one can neglect spatial variation of the distribution function in
the collision volume. Since Eqs.~(\ref{compressGeneral}),(\ref{pcomp}) are
valid for $l_{\rm in}\gg L$, we conclude that the local compressibility is
given by Eqs.~(\ref{compressGeneral})--(\ref{PQ}) provided that $l_{\rm in}\gg
L\gg q_1^{-1}$. The window for $L$ exists if $q_1l_{\rm in}\gg 1$, i.e., if
$\omega_c\tau_{\rm in}\gg 1$, which is satisfied in the
experiments\cite{zudov01,mani02,zudov03,yang03,willett03} by a large margin.

Below we consider the limit of overlapping Landau levels, when the
electron density of states has a weak harmonic modulation
\be
\nu(\ve)=\nu_0\left(1-2\delta\cos{\frac{2\pi \ve}{\omega_c}}\right)\,,\;\;\;\;
\delta=\exp\left(-\frac{\pi}{\omega_c\tau_{\rm q}}\right).
\label{nu}
\ee
From Eq.~(\ref{kineq}) we find:\cite{DVAMP}
\be
\label{fe-expansion}
f_{\rm mw}(\ve)\approx f_0(\ve)+\delta\frac{\omega_{\rm c}}{2\pi}\,
\left(\frac{\partial f_T}{\partial \ve}
\sin \frac{2\pi\ve}{\omega_{\rm c}}\right)F({\cal P},{\cal Q}),
\ee
where the dimensionless function $F$ is given by
\be
\label{distr}
F({\cal P},{\cal Q})=
\frac{{\cal P}\frac{2\pi \omega}{\omega_{\rm c}}
\sin\frac{2\pi\omega}{\omega_{\rm c}}
 +4{\cal Q}} {1+{\cal P}
 \sin^2\frac{\pi\omega}{\omega_{\rm c}}
+{\cal
Q}}.
\ee
Substituting Eq.~(\ref{fe-expansion}) into Eq.~(\ref{pcomp}) for
the compressibility  and assuming (in conformity with the
experiments) that the temperature is not too low,
$2\pi^2 T/\omega_c\gg 1$, we find
\begin{equation}
\delta\chi= -\,\nu_0\,\delta^2\,F({\cal P},{\cal Q}).
\label{compressF}
\end{equation}
Equation (\ref{compressF}) is our main quantitative result. This
equation describes the compressibility of a 2DEG with overlapping
Landau levels, $\delta\lesssim 1$, at arbitrary electric dc and
microwave fields. Most importantly, the compressibility
exhibits oscillations as a function of $\omega/\omega_{\rm c}$.

In the absence of the dc field, the oscillations of $\chi$ have the
same properties as the oscillations of the dissipative part of the
linear conductivity $\sigma_{\rm ph}=\sigma_{\rm
D}[1+2\delta^2(1-F({\cal P},0))]$, see Ref.~\onlinecite{DVAMP}
(here $\sigma_{\rm D}$ is the Drude conductivity). In
particular, the temperature dependence of the amplitude of
oscillations is determined by the inelastic scattering time $\tau_{\rm
in}$.\cite{note2} Note also that at $\omega=k\omega_{\rm c}$ with
integer $k$ the microwave-induced corrections to both the
compressibility and the conductivity vanish.

The above picture of oscillations of the compressibility holds if the
linear conductivity is positive, $\sigma_{\rm ph} >0$. However, if the
dissipative part of the \emph{linear} conductivity is negative, the
electron system becomes unstable and breaks down into
domains.\cite{andreev03} The magnitude of electric field ${\cal
E}_{\rm dc}^*$ inside the domains is set by the condition that the
dissipative electric current $j_{\rm d}=\sigma_{\rm D}{\cal E}_{\rm
dc} [1+2\delta^2(1-F({\cal P},{\cal Q}))]$ is zero. This condition can
be rewritten as $ F({\cal P},{\cal Q}^*)=(1+2\delta^2)/2\delta^2, $
where ${\cal Q}^*$ is expressed in terms of ${\cal E}_{\rm dc}^*$
according to Eq.~(\ref{PQ}). Correspondingly, in the zero-resistance
state the correction to the compressibility is given by
\begin{equation}
\delta\chi^*=-\nu_0\frac{1+2\delta^2}{2},
\label{comprZRS}
\end{equation}
which yields the compressibility for overlapping Landau levels close
to $\nu_0/2$.

These results are illustrated in Fig.~\ref{fig:q1}, where we plot the
compressibility as a function of $\omega_{\rm c}/\omega$ for fixed
values of ${\cal E}_\omega$ and for $\omega\tau_{\rm q}/2\pi=1$.
This choice of $\omega\tau_{\rm q}$ corresponds to typical experimental
values\cite{mani02,zudov03,yang03,willett03} $\omega/2\pi\sim100$~GHz and
$\tau_{\rm q}\sim10$~ps. The oscillatory contribution
$\delta\chi$ is seen to develop with increasing magnetic field,
Eq.~(\ref{compressF}). At sufficiently strong magnetic fields, zero
resistance states appear, domains are formed, and the compressibility
inside the domains is given by $\delta\chi^*$, Eq.~(\ref{comprZRS}).

When crossing a domain wall, the normal component of the dc electric
field changes from $-{\cal E}_{\rm dc}^*$ on one side of the wall to
${\cal E}_{\rm dc}^*$ on the other, thus vanishing in between.  To
estimate the compressibility inside the domain wall, we note that
Eq.~(\ref{compressF}) gives the local compressibility provided the
spatial scale on which the dc electric field varies exceeds the
inelastic relaxation length. The latter gives the characteristic width
of the domain wall, so that we still can use Eq.~(\ref{compressF}) for the
estimate. Then the local compressibility is determined by the strength
of the local electric field ${\cal E}_{\rm dc}$ and $\chi-\chi_0$
changes between
$\delta\chi$, Eq.~(\ref{compressF}), at ${\cal Q}=0$ in the middle of
the domain wall and $\delta\chi^*$, Eq.~(\ref{comprZRS}), away from
the wall, as shown in the inset to Fig.~\ref{fig:q1} for $\omega_{\rm
c}=0.45\omega$.

We recall that Eq.~(\ref{compressF}) is applicable in the regime of
overlapping Landau levels,
$\delta=\exp(-\pi/\omega_c\tau_{\rm q})\ll 1$, when only the
first harmonic $g_1\delta$ of the electron density of states
$\nu(\ve)/\nu_0=1+2\sum^\infty_{l=1}g_l\delta^l\cos(2\pi l\,\ve/\omega_c)$
is important (coefficients $g_l$ are given by\cite{vavilov03}
$g_l=L^1_{l-1}(2\pi l/\omega_c\tau_{\rm q})/l$ and $L^m_l$ are the
Laguerre polinomials). This approximation, Eq.~(\ref{nu}),
works well up to the point
where the Landau levels become separated,
$\omega_c\tau_{\rm q}/2\pi\simeq 1.5$.\cite{vavilov03}
For typical parameters of experiments performed up to date,
$\omega\tau_{\rm q}/2\pi\alt 1$, the approximation is thus sufficient
in the whole range of magnetic fields, $\omega_c<\omega$, where the
microwave-induced oscillations take place.\cite{note-g}
For larger values of $\omega\tau_{\rm q}$, when Landau levels get separated
already at $\omega_c<\omega$, the compressibility as a function of the
magnetic field still has the same qualitative features as
in Fig.~\ref{fig:q1}. In particular, the amplitude of oscillations of
$\delta\chi$ increases as $B$ increases. However, the peaks of
$\delta\chi$ become narrower and plateaus between the peaks appear.
The quantitative analysis is similar to one developed in
Ref.~\onlinecite{DVAMP} and is not presented in this Communication.

The measurement of the local compressibility $\chi$ may be used to
study the strength of the local dc electric fields in the zero
resistance state. Indeed, assuming that the microwave power is
uniformly distributed over the area of the sample, we relate the
variation of the compressibility between different points of a 2DEG to
the variation of the local field ${\cal E}_{\rm dc}$ at these
points. The strength of the field ${\cal E}_{\rm dc}$ can be found
from the measured value of $\chi$ by using Eqs.~(\ref{PQ}) and
(\ref{compressF}). As the electric field decreases from ${\cal E}_{\rm dc}^*$
inside the domains to zero in the middle of the domain wall, the
magnitude of the microwave-induced correction to the
compressibility $|\delta\chi|$ increases and
reaches a maximum at ${\cal E}_{\rm dc}=0$.  Thus, the local minima of
the compressibility $\chi$ mark the boundaries between the domains of
the electric field in the zero resistance state of a 2DEG under
microwave radiation. The length scale that determines the shape of the
minima represents the width of the domain walls.

Finally, we discuss the compressibility in a non-equilibrium state of 2DEG
for an arbitrary value of $ql_{\rm in}$. In the presence of
electrostatic potential $\phi ({\bm r})$, the term
$e\vec {\cal E}_{\rm dc}\partial_\varepsilon$ representing the effect of uniform dc
electric field on electron distribution function acquires a more
general form $e[\nabla\phi({\bm r})-\vec {\cal E}_{\rm dc}] \partial_\varepsilon+\nabla$,
see Ref.~\onlinecite{vavilov03}.
Correspondingly, the second term in the left-hand side of Eq.~(\ref{kineq})
has to be rewritten as
\begin{equation}
\begin{split}
&
\frac{{\cal Q}}{4\nu_0 {\nu}(\ve)}
\frac{\omega_{\rm c}^2 }{\pi^2}\,\partial_\ve
\left[\nu^2(\ve)\partial_\ve f_{\rm mw} \,\right] \to
\frac{D_{\rm c}\tau_{\rm in}}{\nu_0\nu(\varepsilon)} \\
&\times
(e[\nabla\phi-\vec{\cal E}_{\rm dc}] \partial_\varepsilon+\nabla)
\left[
\nu^2(\varepsilon)(e[\nabla\phi-\vec{\cal E}_{\rm dc}]
\partial_\varepsilon+\nabla)
f_{\rm mw}
\right],
\end{split}
\label{subst}
\end{equation}
with $f_{\rm mw}= f_{\rm mw}(\varepsilon,{\bm r})$ and $D_{\rm c}=v_F^2/2\omega_c^2\tau_{\rm tr}$. The substitution
of Eq.~(\ref{subst}) ensures that function $f_0(\varepsilon-e\phi({\bm r}))$
is a solution of kinetic equation
for arbitrary electrostatic potential $\phi({\bm r})$.

The compressibility $\chi$ is related to a linear-in-$\phi$ term
in the solution of the resulting kinetic equation by
$\chi=(\partial/e\,\partial\phi_q)\int d\varepsilon \,\nu
(\varepsilon)\, f_{{\rm mw},q}(\varepsilon)|_{\phi_q\to 0}$, where
$f_{{\rm mw},q}(\varepsilon)$ is the Fourier transform of $f_{\rm
mw}(\varepsilon, {\bm r})$.  Solving the kinetic equation along
the lines of Ref.~\onlinecite{DVAMP} for $q\ll q_1$ (which allows
us to write the inelastic collision integral in the local form in
real space), we obtain for the correction to the compressibility
related to the oscillations of $f_{{\rm mw},q}(\varepsilon)$ in
$\varepsilon$:
\be
\chi-\chi_0=\frac{q^2}{ q^2+l_{\rm in}^{-2}}\,\delta\chi,
\label{compressFanyq}
\ee
where
$\delta\chi$ is given by Eq.~(\ref{compressF}). We note the appearance
of
$l_{\rm in}^{-2}$ term in the denominator of Eq.~(\ref{compressFanyq}).
This term originates from the relaxation of the non-equilibrium
electron distribution function $f_{\rm mw}(\ve)$ due to inelastic
scattering, which occurs on length $l_{\rm in}=(D_c\tau_{\rm
in})^{1/2}$. As a result of such relaxation, the non-equilibrium component
of the  compressibility at $q \ll l_{\rm in}^{-1}$ is suppressed.
However the
compressibility in equilibrium is given by Eq.~(\ref{compressGeneral}) even for
$q = 0$ because a shift of the chemical potential
does not lead to the relaxation of the Fermi distribution function
($l_{\rm in}^{-1}=0$).

We conclude that  the local
relation $\delta n({\bm r})=\chi_{\rm loc}\,e\phi ({\bm r})$ holds only if the
scale $L$ on which $\phi({\bm r})$ changes satisfies the condition $L\ll
l_{\rm in}$, otherwise the oscillatory part of $\delta n({\bm r})$ is
suppressed as $l^2_{\rm in}/L^2$. Therefore, the most favorable conditions for
measuring the microwave-induced oscillations of the local compressibility by
scan probe microscopy should be realized when the size of the probe and the
distance between it and the 2DEG are smaller than or comparable with $l_{\rm
in}$. The technique used in Ref.~\onlinecite{ilani00} makes these conditions
feasible: $l_{\rm in}$ is estimated\cite{DVAMP} to be in the micrometer range
under the experimental
conditions,\cite{zudov01,mani02,zudov03,yang03,willett03} whereas the spatial
resolution in the local compressibility measurements\cite{ilani00} was in the
range of a few tenths of a micrometer.

To summarize, we have shown that the local compressibility $\chi_{\rm loc}$ of
a 2DEG exhibits oscillations governed by the ratio $\omega/\omega_c$,
similarly to the oscillations of the photoconductivity $\sigma_{\rm ph}$. The
key features of the effect of microwave radiation on the compressibility are:
(i) the period and the phase of the oscillations of $\chi_{\rm loc}$ are the
same as for the oscillations of $\sigma_{\rm ph}$; (ii) the amplitude of the
oscillations in $\chi_{\rm loc}$ and $\sigma_{\rm ph}$ have the same
dependence on the electron temperature and microwave power; (iii) at
$\omega=k\omega_{\rm c}$ with integer $k$ the microwave radiation affects
neither $\sigma_{\rm ph}$ nor $\chi_{\rm loc}$; (iv) the zero resistance state
corresponds to a plateau in the compressibility. Local measurements of the
compressibility may be used to make a real-space snapshot of the domain
structure in the zero resistance state.

We thank K.~von~Klitzing, R.~Mani, J.H.~Smet, S.A.~Studenikin, and A.~Yacoby
for discussions of experiments relevant to the
research reported here.
This work was supported by
by the MRSEC Program of the National Science Foundation under
award DMR 02-13282, AFOSR grant F49620-01-1-0457,
and the SPP ``Quanten-Hall-Systeme'' of the DFG.


\begin{thebibliography}{20}
\bibitem[{not({\natexlab{a}})}]{note-sigma}
\bibinfo{howpublished}{In this respect the compressibility is different from
  the conductivity, which contains the factor $\nu^2(\ve)$ in the integrand,
  $\sigma\propto-\int\!d\ve\,\nu^2(\ve)\,\partial f/\partial\ve$. The
  correction to $\sigma$ due to the effect of microwaves on
  $\nu(\ve)$,\cite{ryzhii,durst03,vavilov03} which is smaller by the factor
  $\tau_{\rm q}/\tau_{\rm in}$ than the correction related to the change of
  $f(\ve)$,\cite{dmitriev03, DVAMP} survives at high temperatures, $2\pi^2
  T\gg\omega_c$.}
\bibitem{q_1} In fact, $q_1$ is the smallest characteristic momentum transfer
since the integration over momenta in $\tau_{\rm in}^{-1}$ yields $\ln
(\kappa/q_1)$, where $\kappa\gg q_1$ is the inverse static screening length,
i.e., all momenta between $q_1$ and $\kappa$ contribute ``logarithmically" to
the relaxation rate.
\bibitem[{not({\natexlab{b}})}]{note2}
\bibinfo{howpublished}{The inelastic time $\tau_{\rm in}$ due to the
  electron--electron scattering can be estimated as $\tau_{\rm in} \propto
  E_{\rm F}/T^2$, where $E_{\rm F}$ is the Fermi energy of the 2DEG, see Ref.~\onlinecite{DVAMP}}
\end{thebibliography}

\end{document}